\documentclass[iop]{emulateapj}
\usepackage{epsfig}
\usepackage{apjfonts}
\usepackage{amsmath}

\begin{document}
\title{Interpretation of the unprecedentedly long-lived high-energy emission of GRB 130427A }

\author{Ruo-Yu Liu$^{1,2,4,*}$,  Xiang-Yu Wang$^{1,4}$ and Xue-Feng Wu$^{3,5,6}$}
\affil{$^1$School of Astronomy and Space Science, Nanjing University, Nanjing, 210093, China\\
  $^2$Max-Planck-Institut f\"ur Kernphysik, 69117 Heidelberg, Germany\\
  $^3$Purple Mountain Observatory, Chinese Academy of Sciences, Nanjing 210008, China\\
  $^4$Key laboratory of Modern Astronomy and Astrophysics (Nanjing University), Ministry of Education, Nanjing 210093, China\\
  $^5$Chinese Center for Antarctic Astronomy, Nanjing 210008, China\\
  $^6$Joint Center for Particle, Nuclear Physics and Cosmology, Nanjing University-Purple Mountain Observatory, Nanjing 210008, China\\
  $^*$Fellow of the International Max Planck Research School for Astronomy and Cosmic Physics at the University of Heidelberg (IMPRS-HD)\\}

\begin{abstract}
High energy photons ($>100$ MeV) are detected by the
\textit{Fermi}/LAT from GRB~130427A up to almost  one day after
the burst, with an extra hard spectral component being discovered
in the high-energy afterglow.  We show that this hard spectral
component arises from afterglow synchrotron-self Compton emission.
This scenario can  explain the origin of $>10\,$GeV photons
detected up to $\sim 30000$ s after the burst, which would be
difficult to be explained by synchrotron radiation due to the limited
maximum synchrotron photon energy. The lower energy
multi-wavelength afterglow data can be fitted simultaneously by
the afterglow synchrotron emission. The implication of detecting
the SSC emission for the circumburst environment is discussed.
\end{abstract}


\section{Introduction}
The extended high-energy emission detected by Fermi/LAT is widely
believed to arise from the electrons accelerated in the
external--forward shock via the synchrotron radiation
\citep[e.g.][]{Kumar09, Kumar10}. However, the maximum  photon energy in
this scenario is limited to be $\varepsilon_{\gamma,\rm
max}=50\,\Gamma\,$MeV, where $50$ MeV is the maximum synchrotron
photon in the rest frame of the shock and $\Gamma$ is the bulk
Lorentz factor of the shock, which is usually $\la 300$ at $\sim
100$s after the trigger  \citep[e.g.][]{Piran10,BD11}. So the
detection of $>10\,$GeV photons after 100 s poses a challenge for
the synchrotron emission scenario \citep{Piran10, Sagi12}.
Considering the above difficulty, \citet{Wang13} recently proposed
that the afterglow synchrotron self-Compton (SSC) emission or
external inverse-Compton of central X-ray emission \citep{Wang06}
is likely responsible for the late-time $>10\,$GeV photons seen in
some LAT GRBs.

GRB 130427A triggered the \textit{Fermi}/GBM with a fluence of
$2\times 10^{−3}$erg\,cm$^{-2}$ in 10-1000 keV within a duration
of $T_{90}=138$s \citep{vonKienlin13}. It was later localized  at
$z\simeq 0.34$ \citep{Flores13,Levan13,Xu13a}, implying an
isotropic energy of $7.8\times 10^{53}\,$erg \citep{Kann13}.
Unprecedentedly, $>100\,$MeV emissions are detected well beyond
the prompt emission phase up to about one day after the burst,
including fifteen $>10\,$GeV photons  detected up to $\sim 30000
$s and one 95.3\,GeV photon arrived at 243s after the trigger
\citep{Zhu13,Tam13}. It was  suggested that these  late-time
high-energy photons may arise from inverse-Compton processes
\citep{Fan13, Wang13}. Further evidence comes from the presence of
a hard spectral component (the photon index $\Gamma_{\rm ph}=-1.4\pm 0.1$) above
$2.5\pm 1.1\,$GeV \citep{Tam13} in the late afterglow, which  has
signatures well consistent with the prediction of the afterglow
synchrotron self-Compton emission \citep{Zhang01,Sari01,Zou09}. In this
Letter, we will verify this possibility by modeling the
multi-band (from radio to GeV bands) data of this burst.
Hereafter, we denote by $Q_x$ the value of the quantity $Q$ in
units of $10^x$.

\section{A brief overview of the model}
In the standard synchrotron afterglow spectrum,  there are three
break frequencies, i.e.  $\nu_a$, $\nu_m$ and $\nu_c$, which are
caused by synchrotron self-absorption (SSA), electron injection
and electron cooling respectively. According to
\citep[e.g.][]{Sari98,Wijers99}, these three characteristic
frequencies are given by

\begin{equation}\label{num}
\nu_{{\rm m}} =1.1\times
10^{18}\,f^{2}(p)\epsilon_{e,-1}^{2}E_{54}^{1/2}\epsilon_{{\rm
B},-5}^{1/2}(1+z)^{1/2}T^{-3/2} {\rm Hz},
\end{equation}
\begin{equation}\label{nuc}
\nu_{{\rm c}} =1.1\times 10^{17}\,E_{54}^{-1/2}n_0^{-1}
\epsilon_{{\rm
B},-5}^{-3/2}\left(\frac{1+Y}{100}\right)^{-2}(1+z)^{-1/2}T^{-1/2}
{\rm Hz},
\end{equation}
and
\begin{equation}
\nu_a= 2.7\times 10^{10}\,
f^{-1}(p)E_{54}^{1/5}\epsilon_{e,-1}^{-1}\epsilon_{B,-5}^{1/5}n_{0}^{3/5}(1+z)^{-1},
\end{equation}
in the case that $\nu_a<\nu_m<\nu_c$, which is usually true under
typical parameter values. In the above three equations, $T$ is the
time in the observer's frame since the trigger time,  $z$ is the
redshift of the burst, $f(p)=6(p-2)/(p-1)$ with $p$ being the
electron index, $E$ is the isotropic kinetic energy of the GRB
outflow and $n$ is the number density of the circumburst medium.
For an interstellar medium (ISM) circumburst environment, $n$ is a
constant while for a stellar wind circumburst environment, it is
inversely proportional to the square of the shock radius.
$\epsilon_{{\rm e}}$ and $\epsilon_{{\rm B}}$ are the
equipartition factors for the energy in electrons and magnetic
field in the shock respectively. In the expression of $\nu_c$, $Y$
is the Compton parameter evaluating the effect of SSC cooling on
the synchrotron spectrum. { In the Thomson scattering limit, $Y=Y_0\simeq
(\frac{\epsilon_e}{\epsilon_B})^{\frac{1}{4-p}}\left(\frac{\nu_m}{\nu_{c,\rm
syn}}\right)^{\frac{p-2}{2(4-p)}}$ \citep{Sari01} if $Y\gg 1$},
where $\nu_{c,\rm syn}$ is obtained by considering only the synchrotron cooling.

The time--integrated spectrum of high--energy emission from
GRB~130427A can be modeled by a broken power-law with a soft
component ($\Gamma_{\rm ph}=2.3\pm 0.1$) below the break energy
$E_b=2.5\pm 1.1\,$GeV and a hard component ($\Gamma_{\rm
ph}=1.4\pm 0.1$) above $E_b$ \citep{Tam13}. It is natural to
assume that synchrotron emission is dominant below $E_b$ while the
SSC emission is dominant above $E_b$. As  \citet{Tam13}, we assume
$p=2.2$ in the following analysis. Thus, the synchrotron flux at
$h\nu_c<100 {\rm MeV}\leq h\nu_{\rm LAT}<E_b$ is given by
\begin{equation}\label{100MeV}
\begin{split}
F_{\nu}(100\,{\rm MeV})=& 7.0\, E_{54}^{1.05}\epsilon_{e,-1}^{1.2}\epsilon_{B,-5}^{0.05}(1+z)^{1.05}\\
&\times D_{L,28}^{-2}T^{-1.15}\left(\frac{h\nu}{100\,\rm MeV}\right)^{-1.1}  \mu \rm Jy
\end{split}
\end{equation}
where $D_{L}$ is the luminosity distance. Here we neglect the
inverse-Compton cooling for the electrons that radiate high-energy
gamma-ray emission due to the deep Klein--Nishina (KN) scattering
effect\citep{Wang10, Liu11}. The emission above $E_b$ can be
interpreted as the SSC emission in the regime of $h\nu_m^{\rm
ssc}<h\nu_{\rm LAT}<h\nu_c^{\rm ssc}$, where $\nu_m^{\rm ssc}$ and
$\nu_c^{\rm ssc}$ are the corresponding break frequencies in the
SSC spectrum. The SSC flux is given by \citep{Sari01}
\begin{equation}\label{10GeV}
\begin{split}
F_{\nu}(10{\rm GeV})=&5.6\times 10^{-3} E_{54}^{1.7}n_0^{1.1}\epsilon_{e,-1}^{2.4}\epsilon_{B,-5}^{0.8}(1+z)^{1.5}\\
&\times D_{L,28}^{-2}T^{-1.1}\left(\frac{h\nu}{10\,\rm GeV}\right)^{-0.6} \mu \rm Jy
\end{split}
\end{equation}

Lower energy emission is produced by the forward shock synchrotron
radiation. X-ray observation frequency is usually also in the fast
cooling regime, i.e. $\nu_m<\nu_c<\nu_X$. But the KN suppression
effect is not as important  as in the case of $>100\,$MeV
emission, so we need to consider the effect of SSC cooling on the
synchrotron spectrum carefully. Thus, we have
\begin{equation}\label{X_ray}
\begin{split}
F_{\nu}(10\,\rm keV)=& 17 \,E_{54}^{1.05}\epsilon_{e,-1}^{1.2}\epsilon_{B,-5}^{0.05}\left[\frac{1+Y(10\rm \,keV)}{10}\right]^{-1}\\
&\times(1+z)^{1.05}D_{L,28}^{-2}T^{-1.15}\left(\frac{h\nu}{10\,\rm keV}\right)^{-1.1} \rm mJy
\end{split}
\end{equation}
Here $Y(10{\rm \,keV})=f_{\rm KN}Y_0$ with  $Y_0= U_{\rm syn}/U_B=110\,
E_{54}^{0.056}n_0^{0.056}\epsilon_{e,-1}^{0.67}\epsilon_{B,-5}^{-0.44}(1+z)^{0.056}T^{-0.056}$
being the Compton parameter in the Thomson scattering
limit as mentioned above, where $U_{\rm syn}$ and $U_B$ are the energy density of
synchrotron radiation field and magnetic field respectively. The
factor $f_{\rm KN}$ considers the KN effect on the electrons that
emit $10\,$keV photons by synchrotron radiation. In the Thomson
scattering regime $f_{\rm KN}=1$, while in the deep KN scattering
regime $f_{\rm KN}\rightarrow 0$.

The KN effect will intervene in the inverse Compton cooling if the
energy of the incident photon exceeds the rest energy of the
scattering electron in its rest frame, i.e., $\gamma_e h\nu_{\rm
KN}(1+z)/\Gamma \geq m_e c^2$, where $\gamma_e$ is the Lorentz
factor of the electron in the comoving frame of the emitting
region and $\nu_{\rm KN}$ is the critical frequency of the
incidence photon measured in the observer's frame. If the energy
of a photon is larger than $h\nu_{\rm KN}$, the scatter will enter
the KN regime with a suppressed cross section. The Lorentz factor
of the electron emitting X-ray photons can be given by
$\gamma_{\rm 10\,keV}=\left( \frac{2\pi m_e c \nu_{\rm X}(1+z)}{\Gamma
eB} \right)^{1/2}$ and thus
\begin{equation}\label{nuKN}
\nu_{\rm KN, 10\,keV}=2.8\times 10^{18}\,E_{54}^{1/4}\epsilon_{B,-5}^{1/4}(1+z)^{-3/4}\left(\frac{h\nu_{\rm X}}{10\,\rm keV}\right)^{1/2}T^{-3/4}
\end{equation}
As $\nu_{\rm KN}\propto T^{-3/4}$,  the KN suppression effect
becomes more and more important at later time. As a rough
estimate, $f_{\rm KN}$ can be given by
\begin{equation}\label{fKN}
f_{\rm KN}=\frac{U_{\rm syn}(\nu<\nu_{\rm KN})}{U_{\rm syn}}\simeq \left\{
\begin{array}{lll}
0.06\left(\frac{\nu_m}{\nu_c}\right)^{0.4}\left(\frac{\nu_{\rm KN}}{\nu_m}\right)^{4/3} & \nu_{\rm KN} < \nu_m,\\
0.2\left(\frac{\nu_{\rm KN}}{\nu_c}\right)^{0.4} & \nu_m < \nu_{\rm KN} < \nu_c,\\
1-0.8\left(\frac{\nu_{\rm KN}}{\nu_c}\right)^{-0.1} & \nu_c < \nu_{\rm KN}
\end{array}
\right.
\end{equation}
In a broad parameter space, we find
$\nu_m < \nu_{\rm KN} <\nu_c$, leading to $f_{\rm KN}\simeq 0.1
E_{54}^{0.3}n_{0}^{0.4}\epsilon_{B,-5}^{0.7}(1+z)^{0.1}[\frac{1+Y(10\,{\rm
keV})}{10}]^{0.8}(\frac{h\nu}{10\rm keV})^{0.2}T^{-0.1}$.

Optical emission is typically in the frequency regime
$\nu_m<\nu_{\rm Op}<\nu_c$, so the  flux is given by
\begin{equation}\label{opt}
\begin{split}
F_{\nu}({\rm R\,band})=& 1.3\, E_{54}^{1.3}n_0^{0.5}\epsilon_{e,-1}^{1.2}\epsilon_{B,-5}^{0.8}(1+z)^{1.3}\\
&\times D_{L,28}^{-2}T^{-0.9}\left(\frac{\nu}{4.56\times
10^{14}\rm Hz}\right)^{-0.6} \rm Jy.
\end{split}
\end{equation}
The radio observation frequency could  lie in the regime
$\nu_a<\nu_{\rm Ra}<\nu_m$ or $\nu_{\rm Ra}<\nu_a<\nu_m$. In the
former case,
\begin{equation}\label{radio1}
\begin{split}
F_{\nu}(5\,{\rm GHz})=&20\, E_{54}^{5/6}n_0^{1/2}\epsilon_{e,-1}^{-2/3}\epsilon_{B,-5}^{1/3}(1+z)^{5/6}\\
&\times D_{L,28}^{-2}T^{1/2}\left(\frac{\nu}{5\,\rm
GHz}\right)^{1/3} \mu \rm Jy,
\end{split}
\end{equation}
while in the latter case,
\begin{equation}\label{radio2}
\begin{split}
F_{\nu}(5\,{\rm GHz})=&1.3\, E_{54}^{1/2}n_0^{-1/2}\epsilon_{e,-1}(1+z)^{5/2}\\
&\times D_{L,28}^{-2}T^{1/2}\left(\frac{\nu}{5\,\rm GHz}\right)^2
\mu \rm Jy.
\end{split}
\end{equation}

We note that in previous works, modeling of the multi-band light
curves of LAT-detected GRBs usually result in a low circumburst
density $<10^{-2}\,\rm cm^{-3}$ \citep[e.g.][]{Cenko11, Liu11,
He11}. In a low-density environment, SSC flux would be severely
suppressed. We here point out that the low-density result is
mainly due to neglect of the KN effect in modeling
the X-ray afterglows. We now briefly show that considering the KN
effect in X-ray afterglow can change the inferred density
dramatically.

Usually $>10\,$GeV data are not available, and only $>100\,$MeV,
X-ray, optical and radio data are used in the multi-band light
curve fit. So from Eqs.~(\ref{100MeV}), (\ref{X_ray}),
(\ref{opt}), and (\ref{radio1}) or (\ref{radio2}), we get four
independent constraints  on four undetermined parameters, namely
$E$, $n$, $\epsilon_e$ and $\epsilon_B$ (note that $p=2.2$ is
fixed) . So we can fully determine these parameters by solving
these equations. In the case that KN effect is not considered in
X-ray emission, i.e., $f_{\rm KN}=1$, we get
\begin{equation}
\left\{
\begin{array}{llll}
E\simeq C_{\rm LAT}^{0.74}C_{\rm X}^{0.45}C_{\rm Op}^{-0.59}C_{\rm Ra}^{0.64}\\
n\simeq C_{\rm LAT}^{0.28}C_{\rm X}^{-2.26}C_{\rm Op}^{1.09}C_{\rm Ra}^{0.66}\\
\epsilon_{e}\simeq C_{\rm LAT}^{0.26}C_{\rm X}^{-0.45}C_{\rm Op}^{0.48}C_{\rm Ra}^{-0.53}\\
\epsilon_{B}\simeq C_{\rm LAT}^{-1.77}C_{\rm X}^{1.35}C_{\rm Op}^{0.80}C_{\rm Ra}^{-0.65}.
\end{array}
\right.
\end{equation}
Here we adopt the case that $\nu_a<\nu<\nu_m$ for radio flux as an
example.  $C_{\rm LAT}$, $C_{\rm X}$, $C_{\rm Op}$ and $C_{\rm
Ra}$ are constants related to data used to normalize the flux in
LAT, X-ray, optical, radio band respectively.

When considering the KN effect in X-ray, as long as $Y=f_{\rm
KN}Y_0\gtrsim 1$, we  can approximately replace the $C_{\rm X}$ by
$C_{\rm X}f_{\rm KN}$ in the above equation set and we then get
\begin{equation}\label{KN_noKN}
\left \{
\begin{array}{llll}
E=f_{\rm KN}^{0.45}\hat{E}\\
n=f_{\rm KN}^{-2.26}\hat{n}\\
\epsilon_{e}=f_{\rm KN}^{-0.45}\hat{\epsilon}_{e}\\
\epsilon_{B}=f_{\rm KN}^{1.35}\hat{\epsilon}_{B}.
\end{array}
\right.
\end{equation}
Here we denote the parameters without considering  the KN effect
on X-ray emission by  hatted characters. If $f_{\rm KN}Y_0\ll 1$,
$f_{\rm KN}$ should be replaced by $(1+Y_{0})^{-1}$ in the above
equation set. Since $f_{\rm KN}<1$, considering the KN effect in
X-ray emission will increases the inferred ISM density
significantly. Substituting Eq.~(\ref{KN_noKN}) into
Eq.~(\ref{10GeV}), we find $F_{\nu, \rm IC}\simeq f_{\rm
KN}^{-1.7}\hat{F}_{\nu, \rm IC}$, so the SSC flux can be
significantly increased as well.

{Here we would like to point out that Eq.~(\ref{fKN})
underestimates the  value of $f_{\rm KN}$ because the scattering
cross section in the KN regime is approximated as being zero. In
our numerical code, we calculate $f_{\rm KN}$ as
\begin{equation}
f_{\rm KN}=\frac{\int_0^{\nu_{\rm KN}}F_\nu d\nu + \int_{\nu_{\rm KN}}^{\infty}
 [2{\rm ln}\,(2\gamma_e h\nu/\Gamma m_ec^2)+1](\frac{\nu}{\nu_{\rm KN}})^{-1}F_\nu d\nu}{\int_0^{\infty}F_\nu
 d\nu},
\end{equation}
where the expression before $F_\nu$ in the second integration on
the numerator accounts for the correction to the scattering cross
section above $\nu_{\rm KN}$.}

\section{Fitting of the multi-band light curve data }
There are abundant observational data of the multi-band light
curves of GRB~130427A.  The extended high-energy emission in the
energy range 0.1-2\, GeV and  2-100\,GeV  shows a power-law decay
with slopes of $\alpha_1=−1.1\pm 0.1$ and $\alpha_2=−1.0\pm
0.1$ respectively \citep{Tam13}. Therefore we adopt the ISM
density profile for the circumburst environment, because in the
wind environment, the late-time SSC flux would decrease much
faster than the observed one \footnote{In the ISM medium, $\nu_{m,
\rm IC}\propto T^{-9/4}$, $F_{\nu\rm m, IC}\propto T^{1/4}$ and
$F_{\nu}=F_{\nu\rm m, IC}(\nu/\nu_m)^{-(p-1)/2}\propto
T^{(11-9p)/8}$. While in the wind medium, $\nu_{m,\rm IC}\propto
t^{-2}$ and $F_{\nu\rm m, IC}\propto t^{-1}$, so $F_{\nu}=F_{\nu
\rm m, IC}(\nu/\nu_m)^{-(p-1)/2}\propto t^{-p}$}.

According to the Fermi/LAT spectral data shown in  \citep{Tam13},
the 0.1-2\,GeV flux is dominated by the synchrotron component.
Requiring the synchrotron flux be $10^{-4}\,$ph\,cm$^{-2}$s$^{-1}$
at 300s, we obtain
\begin{equation}\label{C_100MeV}
E_{54}^{1.05}\epsilon_{e,-1}^{1.2}\epsilon_{B,-5}^{0.05}=1.5.
\end{equation}
And, by requiring the SSC flux  be $3\times
10^{-7}\,$ph\,cm$^{-2}$s$^{-1}$ at 5000s in 2-100\,GeV, we obtain
\begin{equation}
E_{54}^{1.7}\epsilon_{e,-1}^{2.4}\epsilon_{B,-5}^{0.8}n_0^{1.1}=20.
\end{equation}

The XRT data show that the X-ray flux decay with a slope of -1.2
since 421s  after the trigger, then breaks at 53.4\,ks to a
steeper slope of -1.8 \citep{Evans13}, and after $\sim 1\,$day the
light curve becomes shallower again with a slope of -1.3. The
first slope is consistent with the decay of the synchrotron flux
when the observation frequency is above the cooling frequency,
i.e. $\nu_{\rm X} >\nu_c$. To explain the second slope, we
introduce a jet break around 53.4\,ks. The late-time shallower
decay could be due to  the KN suppression effect which becomes
more and more important at later times. Since the KN effect is not
easy to express accurately in an analytical way, we include it in
the numerical modeling.

The optical flux decreases slowly at early time with a slope of
-0.8\citep{Laskar13},  which is consistent with the decay slope of
synchrotron emission in the frequency regime $\nu_m <\nu_{\rm opt}
<\nu_c$. The light curve becomes obviously steeper after $\sim
0.3\,$day  with a slope of -1.35 \citep{Laskar13} and then a
flattening shows up at $\sim 10\,$days after the
trigger\citep{Trotter13}. The steepening in the light curve can be
ascribed to the jet break as well. Although the jet break usually
results in a steeper slope than -1.35, we note that there is an
emerging supernova component \citep{Xu13b}, so this shallow slope
could be caused by the superposition of the fast decaying
synchrotron afterglow component and the supernova component. Since
the R-band flux is about 1\,mJy at $\sim30$\,ks, we have
\begin{equation}
E_{54}^{1.3}\epsilon_{e,-1}^{1.2}\epsilon_{B,-5}^{0.8}n_{0}^{0.5}=1.8.
\end{equation}

The radio data starts from $\sim 0.67\,$day after the trigger
\citep{Laskar13},  which is around the assumed jet break time. The
observed 5\,GHz flux at  0.67\,day and 2\,days after the trigger
are comparable, and decreases by a factor of a few  at 4.7\,days.
This  decay can be ascribed to the jet break. Higher frequency
observations such as $20\,$GHz, $36\,$GHz start later and show a
decaying light curve from the beginning. By requiring  that the
synchrotron flux at 5\,GHz flux be 2mJy at 0.67\,day, we get
\begin{equation}
E_{54}^{0.5}\epsilon_{e,-1}n_0^{-0.5}=0.9.
\end{equation}
The early radio emission can be also attributed to the reverse
shock emission, as suggested by
\citet{Laskar13}\footnote{Introducing a reverse shock component
helps to explain the observed soft radio spectrum ($F_{\nu}\propto
\nu^{\beta}$ with $\beta<0$) at 2\,days and 4.7\,days after the
trigger time. On the other hand, we also note that radio
observations during early time (e.g., <10\,days after the trigger
time) may suffer from the interstellar scintillation
\citep{Goodman97}, so the observed flux and spectral index could
be affected to some extent.}. Then we will require that the flux
produced by the forward shock is below the observed flux. This is
possible in the framework of the decaying micro-turbulence
magnetic field scenario \citep{Lemoine13, Wang13}, since the
radio-emitting electrons radiate in a weaker magnetic field
region.

As we can see, the early and late time  LAT observations, the
optical and the radio observations already provide four
independent constraints on four parameters (i.e., $E$, $n$,
$\epsilon_e$ and $\epsilon_{B}$). Since the X-ray observation can
in principle put another constraint independently, the system of
equations is overdetermined. Finding a solution for the
overdetermined system of equations supports the validity of our
model.

\begin{figure}[]
\epsscale{1.2} \plotone{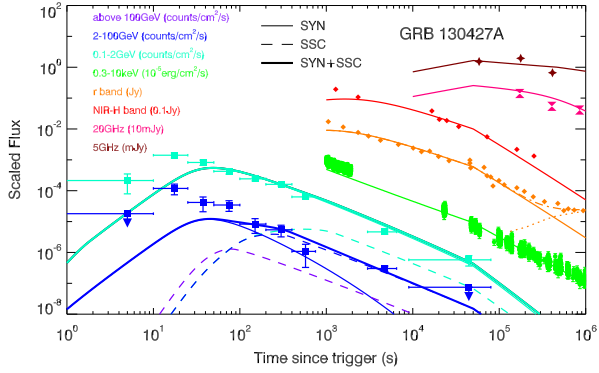}\caption{Fit of the
multi-band light curves  of GRB~130427A. For the high-energy
emission, the thin solid lines represent the synchrotron flux, the
dashed lines represent the SSC flux and the thick solid lines
represents the total flux. The orange dotted line represents the
supernova component, which has the same shape as SN1998bw but with
a flux normalized to the detected flux. And the orange dotted
line is the sum of the GRB afterglow component and the supernova component. The LAT data is
taken from \citep{Tam13}. The X-ray data is taken from the
website http://www.swift.ac.uk/xrt\_curves. The optical,
near-infrared data and  radio data are taken from \citep{Laskar13}.
In this fit, we used the following parameter values: $E_k=
2\times 10^{53}\,$erg, $\epsilon_e=0.6$, $\epsilon_B= 1.3\times
10^{-5}, n= 1\,\rm cm^{-3}$, $p=2.2$, $\Gamma_0=200$. The jet
break time is set at 0.65\,day after the trigger time. The opening
angle is $\theta_j= 7^{\circ}$, corresponding to $E_{k,\rm jet}=
1.5\times 10^{51}\,$erg. \label{lc}}
\end{figure}

\begin{figure}[]
\epsscale{1.2} \plotone{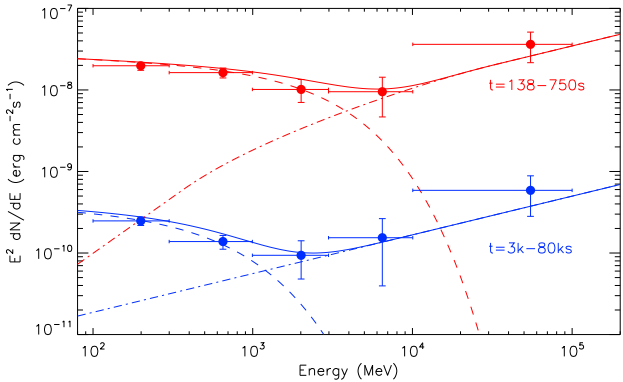}\epsscale{1.2}\caption{Fit of
the spectrum of LAT emission at two different times. The
dashed lines and the dash--dotted lines represent the synchrotron
component and the SSC component respectively, and the solid lines
represent the sum of them.\label{spectrum}}
\end{figure}

By solving the above four equations, we  get $E_{54}\simeq 0.3$,
$n_0\simeq 6$, $\epsilon_{e,-1}\simeq 5$, $\epsilon_{B,-5}\simeq
0.5$.
Although the analytic
solution and numerical solution may not conform each other
perfectly, we can use these values as a guide, and fine-tune them
to get a good global solution in our numerical code. We show the
fitting of multi-band light curves in Fig.~\ref{lc}. The final
values of the parameters are not much different from the above
values, as shown in the caption of the figure. We also present the
light curve of $>100\,$GeV SSC emission with a purple dashed line.
It peaks around 100s with a flux of
$10^{-6}$ph\,cm$^{-2}$s$^{-1}$. Given the LAT effective area  of
$\sim 10^4$cm$^{-2}$, one  expect that LAT may detect one $\gtrsim
100\,$GeV photon in $\sim 100$s around the peak time, which is
consistent with the detection of the $95.3\,$GeV photon at 243s
after the trigger time. Our model can not explain the high energy
emission during the prompt emission phase ($\lesssim 100$s), which
is attributed to the internal dissipation origin \citep{He11,
Liu11, Maxham11}. In the numerical modeling, we find that $f_{\rm
KN}(10\,\rm keV)$ gradually decreases from $\gtrsim 0.1$ at
$1000$s to $\sim 0.01$ at $10^{6}$s. Fig.~\ref{spectrum} presents
the fit of the time-integrated spectrum of LAT emission for the
period of $138-750$ s and $3000-80000$ s.

\section{Discussions}
The extended,  hard emission above a few GeV, seen in GRB 130427A,
represents strong evidence of a SSC component in the forward shock
emission. The appearance of the SSC component implies that the
circumburst density should not be too low, in contrast to the
result in the previous study that the circumburst density of
LAT--detected bursts is on average lower than usual
\citep{Cenko11}. The inferred density in this work is of the same
order of the typical ISM density in the galaxy disk where massive
stars reside.

Recently,  \citet{Laskar13} modeled the low-energy afterglows of
this GRB with a forward-reverse shock synchrotron emission model.
However, we find that their model predicts a synchrotron flux
about one order of magnitude lower than the observed flux in the
LAT energy range. To interpret the multi-band afterglow data
including the LAT data, we proposed a forward shock synchrotron
plus SSC emission scenario. We also find that an ISM environment
is favored by the slow decay of 2--100\,GeV flux, which is
explained as the SSC origin.

We would like to thank He Gao, Zhuo Li and the anonymous referee for valuable suggestions.
This work is supported by the 973 program under grant
2009CB824800, the NSFC under grants 11273016,  10973008, and
11033002, the Excellent Youth Foundation of Jiangsu Province
(BK2012011), and the Fok Ying Tung Education Foundation.
XFW is partially supported by the National Basic Research Program ("973" Program) of China (Grant
2013CB834900), the One-Hundred-Talents Program and the Youth Innovation Promotion Association
of the Chinese Academy of Sciences, and the Natural Science Foundation of Jiangsu Province.

\end{document}